\documentclass[review,11pt]{elsarticle}

\usepackage{graphicx}
\usepackage{float}
\usepackage{subfigure}
\usepackage{amssymb}
\usepackage{amsmath}

\usepackage{lineno}

\usepackage{geometry}
\usepackage{color}
\usepackage{multicol}
\usepackage{multirow}
\usepackage{booktabs}
\usepackage{array}
\usepackage{threeparttable}
\usepackage{times} 
\usepackage{siunitx}
\usepackage{pdfpages}
\journal{Nature Computational Science}

\makeatletter
\def\ps@pprintTitle{%
\let\@oddhead\@empty
\let\@evenhead\@empty
\def\@oddfoot{}%
\let\@evenfoot\@oddfoot}
\makeatother

\begin{document}
\renewcommand{\abstractname}{Summary}

\begin{frontmatter}


\title{Multiscale coupling of surface temperature with solid diffusion in large lithium-ion pouch cells}



\author[eng]{Jie Lin}
\author[eng]{Howie N. Chu}
\author[eng,FI]{David A.\ Howey}
\author[eng,FI]{Charles W.\ Monroe\corref{cor1}}
\address[eng]{Department of Engineering Science, University of Oxford, Oxford, OX1 3PJ, United Kingdom}
\address[FI]{The Faraday Institution, Harwell Campus, Didcot, OX11 0RA, United Kingdom}
\cortext[cor1]{Corresponding author}

\begin{abstract}
Untangling the relationship between reactions, mass transfer, and temperature within lithium-ion batteries enables control approaches that mitigate thermal hot spots and slow degradation. Here, we develop an efficient physics-based pouch-cell model to simulate lock-in thermography experiments, which synchronously record the applied current, cell voltage, and surface-temperature distribution. Prior modelling efforts have been confounded by experimental temperature profiles whose characteristics suggest anisotropic heat conduction. Accounting for a multiscale coupling between heat flow and solid-state diffusion rationalizes this surface-temperature nonuniformity. We extend an earlier streamlined model based on the popular Doyle--Fuller--Newman theory, augmented by a local heat balance. The reduced-order model is exploited to parametrize and simulate commercial 20 Ah lithium iron phosphate (LFP) cells at currents up to 80 A. This work highlights how microscopic intercalation processes produce distinctive macroscopic heat signatures in large-format cells, as well as how heat signatures can be exploited to fingerprint material properties.
\end{abstract}


\end{frontmatter}


\section{Introduction}
\label{S:1}
Specific energy, cycle life, safety, and cost of lithium-ion batteries have all substantially improved in the past decade \cite{Gibb2021:rise,Chu2016;Path,Nitta:2015Review,Ciez:2020cost}, but challenges remain for high-power applications. The heat generation that accompanies charge or discharge \cite{Bernardi:1985Ge} generally increases both a cell's mean temperature and the extent of its temperature non-uniformity, which can significantly degrade performance and cycle life \cite{Ma:2018Review,Finegan:2015In,Zhu:2019Fast,Wang:2020Under}. In worst-case scenarios, non-uniform heating of lithium-ion batteries at high power can facilitate catastrophic thermal runaway \cite{Feng:2018Review,Liu:2018Th}. Outside of the effect on degradation, local variations of temperature within a battery cell can significantly impact characteristics such as equilibrium voltage and apparent internal resistance \cite{Yang:2019Asy,Yang:2021The}. Temperature measurements contain rich information about the physicochemical processes that govern battery behaviour. If the microscopic origins of a battery's thermal footprint are understood in sufficient detail, it may be possible to diagnose complex microscopic information from macroscopic temperature measurements. Temperature distributions can be monitored transiently via non-invasive thermal imaging techniques \cite{Robinson:2016Review,Finegan:2017Ch,Howie:2020pa,Wang:2020Th} and implanted sensors \cite{Huang:2020Op}.  

Battery models are essential tools for exploring how different physical mechanisms contribute to measured behaviour \cite{Li:2020Mul,Mistry:2018Ele}. Experimentally validated models are further useful for estimating material properties and optimising cell designs \cite{Deng2018:Safe}. Physics-based electrochemical simulations based on porous-electrode theory, such as the Doyle--Fuller--Newman model \cite{Doyle:1993Mo,Fuller:1994Re,Fuller:1994Si}, are well established. Almost all of the numerous porous-electrode-theory investigations in the literature focus only on the electrochemical response of a single layer in the `through-plane' direction perpendicular to the current collectors \cite{Doyle:1996Ca,Cai2011:Ma,Yang:2017Mo,Yao2019:Quan,Finegan2020:spa,Lu2020:3D}. The `in-plane' distribution of current can be equally or perhaps even more important, however, especially in large-format cells. 

This paper shows that the surface-temperature distribution across a large-format cell is an effective probe for battery diagnostics. We investigate the causes of non-uniform in-plane temperature distributions during battery charging and discharging, and explore various electrochemical processes that may rationalize them. Through use of a judiciously designed test rig that minimizes heating due to external wiring and tab contacts \cite{Howie:2020pa}, lock-in thermography \cite{Robinson:2015De} of commercial large-format 20 Ah LFP/Graphite pouch cells from A123 Systems is performed while the cells undergo square-wave cycling or constant-current discharge. The results of these experiments are simulated using a new computationally efficient three-dimensional battery model. 

We formally derive the streamlined model previously proposed by Chu \emph{et al.} \cite{Howie:2020pa} from a three-dimensional version of the Doyle--Fuller--Newman model that is extended with a local heat balance. This process reveals several natural routes to produce reduced-order models that account rigorously for additional phenomena, while retaining the high computational efficiency and parsimonious parameter set of the streamlined model.

Excellent agreement between simulations and experiments is obtained by extending the streamlined model to include solid-phase diffusion dynamics. A single lumped diffusion time for both electrodes suffices to match experimental temperature profiles. The three-dimensional aspect of the electrode model differs substantially from typical approaches based on porous-electrode theory, and provides new insights into the dominant physical mechanisms that result in non-uniform temperature distributions. As well as reducing the large set of unknown material properties involved in Doyle--Fuller--Newman theory to a set of just a few observable parameters, the order reduction makes our model computationally efficient enough to support inverse-modelling algorithms based on iterative solutions of full finite-element simulations. We demonstrate, by extracting parameter values from measured cell data, that the streamlined model with solid diffusion can accurately estimate solid-phase diffusivity, as well as key material properties such as electrolyte conductivity, interfacial exchange-current density, specific heat capacity, and cell-reaction entropy, among others.

\section{Temperature non-uniformity and solid-state diffusion}
\label{results:nonuniform}

\begin{figure}
\centering\includegraphics[width=13cm]{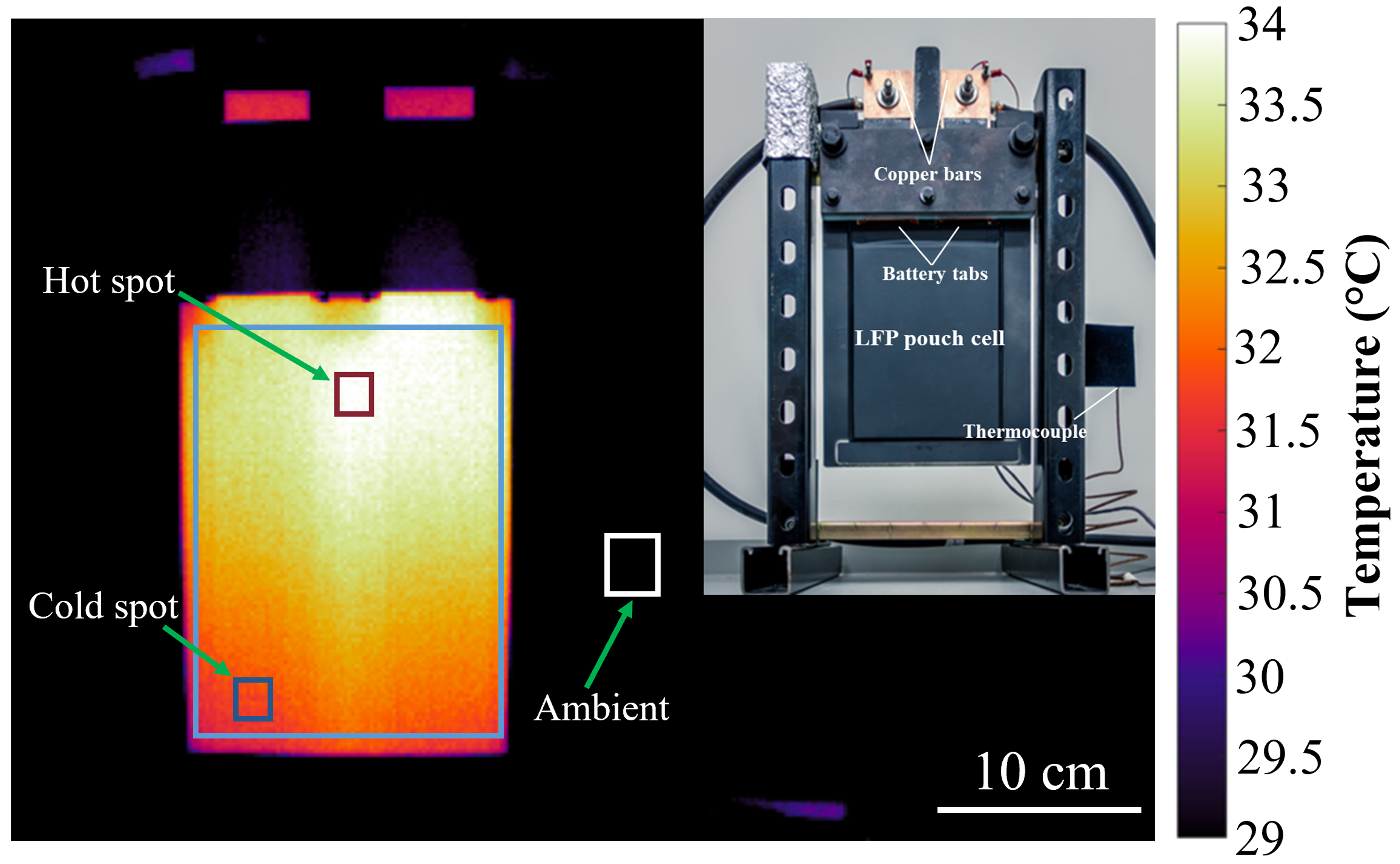}
\caption{An infrared image of the instantaneous surface-temperature distribution across a 20 Ah pouch cell during lock-in thermography. Inset: Photograph of the experimental test rig.}
\label{fig:experimental_rig}
\end{figure}

Lock-in thermography experiments were conducted to measure the surface temperatures of pouch cells synchronously with their voltage output under given applied currents. Experimental data was gathered using the test rig depicted in Fig.\ \ref{fig:experimental_rig}, for which the experimental setup and procedures were established by Chu \emph{et al.}\ \cite{Howie:2020pa}. 
Here we report data from two sources: square-wave-excitation cycling experiments, of which the data sets that cycled around a 30\% state of charge (SOC) were reported earlier \cite{Howie:2020pa}, but the others were not; and full-cell discharge experiments, performed specially for this report. All experiments used 20 Ah pouch cells from A123 Systems, which have a lithium iron phosphate (LFP) positive electrode and a graphite negative electrode. 

In all cases, lock-in thermography was performed using cells initially equilibrated at ambient temperature. Before each cycling experiment, the cell was discharged from 100\% SOC using Coulomb counting to a predetermined initial SOC of 30\%, 50\%, or 70\%. Cells were cycled galvanostatically, alternating between charge and discharge periods of equal length for the 2500 s duration of the experiment, although the first charge step was performed over a half-period to keep the cell's time-averaged SOC centred at its initial value. The applied current was set at 2C or 4C, with periods of 50 s or 100 s for one charge/discharge cycle. The cell voltage and ambient temperature were recorded at 1.0 Hz.

Thermograms of the cell surface were captured via the thermal imaging camera, allowing online visual monitoring of cell temperature. A physics-based battery model was solved using COMSOL Multiphysics software to simulate the electrical and thermal responses to the square-wave current excitation (see Methods section for more details). For all the square-wave cycling tests, parameter estimation based on a previously introduced streamlined model \cite{Howie:2020pa} could capture the hot-spot, cold-spot, and surface-average temperatures accurately, but failed to predict the correct horizontal temperature distribution on the surface --- that is, the distribution across the largest surface of the pouch cells, in the direction perpendicular to the tabs (see Figure\ S2 of the Supplementary Material). 

In hopes of improving fits of the horizontal temperature distribution, several reduced-order models were derived, each based on the Doyle--Fuller--Newman (DFN) model with an added local heat balance. These produced a variety of extensions to the streamlined model of Chu \emph{et al.}\ \cite{Howie:2020pa}, as described in Supplemental Note 2. We found that the assumption of linear kinetics in place of nonlinear Butler--Volmer kinetics did not affect observed surface-temperature distributions (see Figure\ S3). The inclusion of solid-state diffusion, however, did have a fairly large macroscopic effect. In particular, solid-state diffusivity was found to be the only parameter that had observable impact on the horizontal temperature distribution. This finding suggests that there is a close coupling between thermal transport and solid-state diffusion in the electrode particles, and further, that consideration of this coupling is necessary to account for the horizontal temperature variation in large-format pouch cells.

\begin{figure}
\centering  
\includegraphics[width=12cm]{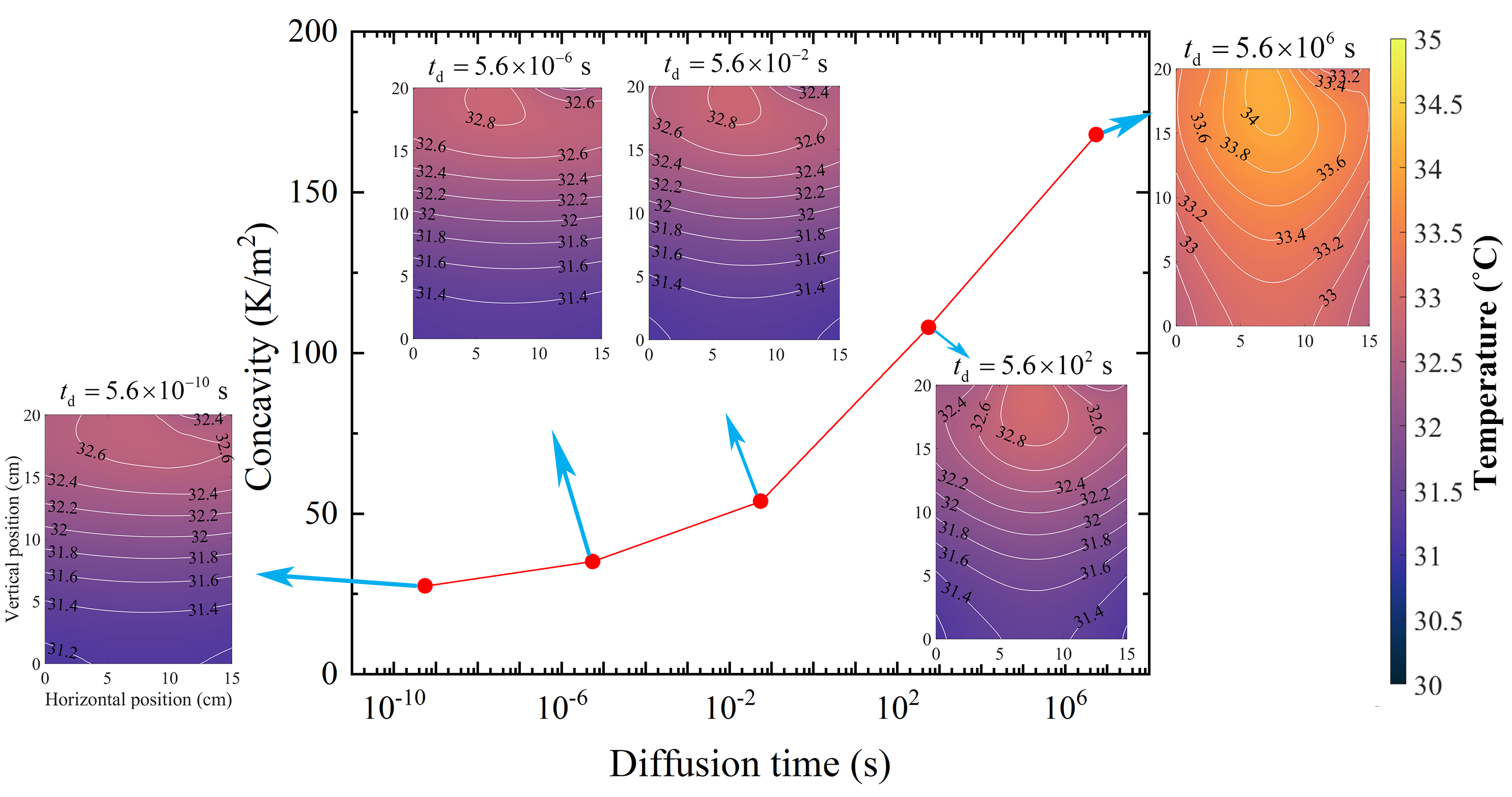}
\caption{Coupling of horizontal temperature concavity through the hot spot with solid-state diffusion time. Insets: simulated surface-temperature profiles after $2500$~s of square-wave cycling with 80 A applied-current amplitude and 100 s period with a 30\% initial state of charge (4C--100s@30\%), at various diffusion times $t_\textrm{d}$.}
\label{Fig.2}
\end{figure}

Figure \ref{Fig.2} presents simulation results that demonstrate how the spatial concavity of the horizontal temperature distribution (along a horizontal axis through the hot spot) varies with the solid-state diffusion time constant $t_\textrm{d}$ (\emph{cf.}\ Table \ref{tab:fitted_values}).
\begin{table}
\centering
\caption{Parameters yielded by model fits based on the new reduced-order model (This work) and the model of Chu \emph{et al.} \cite{Howie:2020pa} (Streamlined model), extracted from experimental lock-in thermography data under 4C--100s square-wave applied currents, with differing initial states of charge.}
\label{tab:fitted_values}
\begin{threeparttable}
\resizebox{\linewidth}{!}{%
\begin{tabular}{l|c|ccc|ccc} 
\hline
\multicolumn{1}{c|}{\multirow{2}{*}{Parameter}}                                               & \multirow{2}{*}{Symbol}                              & \multicolumn{3}{c|}{This work}                              & \multicolumn{3}{c}{Streamlined model}                                                                                  \\ 
\cline{3-8}
  &             & 30\%   & 50\% & 70\%       & 30\%          & 50\%         & 70\%                      \\ 
\hline
Local reaction current density ($\si{A.cm^{-3}}$)    & $ai_0^\theta$       & $1.86$ & $1.80$ & $1.86$        & $143$       & $120$      & $119$            \\
Reaction activation energy ($\si{kJ.mol^{-1}}$)     & $E^\theta$ & $29.5$ & $29.2$ & $29.7$     & $30.8$  & $30.6$             & $28.8$            \\
Effective ionic conductivity ($\si{S.m^{-1}}$)     & $\kappa$      & $0.046$ & $0.043$ & $0.045$      & $0.022$ & $0.019$       & $0.022$   \\
Temperature coefficient of $\kappa$ ($\si{\milli S.m^{-1}.K^{-1})}$)  & $\alpha$ & $2.4$ & $2.7$ & $2.4$   & $2.0$    & $2.0$      & $3.0$   \\
OCP gradient ($\si{V}$)                         & $k_U$         & $0.35$   & $0.24$ & $0.18$         & $0.36$     & $0.22$             & $0.17$       \\
Diffusion time ($\si{s}$)      & $t_\textrm{d} = \frac{r_0^2}{D_s}$          & $552$ & $590$ & $622$ & \multicolumn{3}{c}{--}         \\
Entropy change ($\si{J.mol^{-1}.K^{-1}}$)    & $\Delta S$  & $-13.5$ & $7.7$ & $9.7$     & $-13.5$          & $7.7$              & $9.7$     \\
Heat transfer velocity ($\si{\micro m.s^{-1}}$)      & $\frac{h}{\tilde{C_p}}$      & $5.11$ & $5.06$ & $5.00$  & $5.39$ & $5.20$ & $4.98$  \\
Effective thermal conductivity ($\si{W.m^{-1}.K^{-1}}$)     & $k$  & $1.1$ & $1.2$ & $1.1$       & $54.4$       & $72.0$     & $60.1$   \\
\hline
\end{tabular}
}
\end{threeparttable}
\end{table}
Details of how this concavity was estimated are provided in Supplemental Note 1 and Figure S4. The horizontal concavity flattens as the diffusion time constant decreases, ranging from 168 to 35 \si{K.m^{-2}}. At $t_\textrm{d}=5.6\times10^{-6}$~\si{s} --- an unrealistically fast value \cite{Dees:2020Ap,Chang:2021Im} --- the hot spot lies very close to the cell's top edge, and the horizontal temperature variation is minimal across the vast majority of the cell surface. This behaviour qualitatively agrees with results presented by Chu \emph{et al.}\ \cite{Howie:2020pa}, whose model derives from the assumption that solid-state diffusivity is infinitely large, as explained in Supplemental Note 2. If the diffusion time constant increases (\emph{i.e.}, the diffusion coefficient decreases), then horizontal temperature non-uniformity increases. At diffusion time constants above \SI{100}{s}, significant extra heat generation occurs, causing an even larger temperature difference between the central vertical axis of the cell surface and its left or right edges, with generally higher absolute temperatures everywhere.  

\begin{figure}
\centering  
\includegraphics[width=13cm]{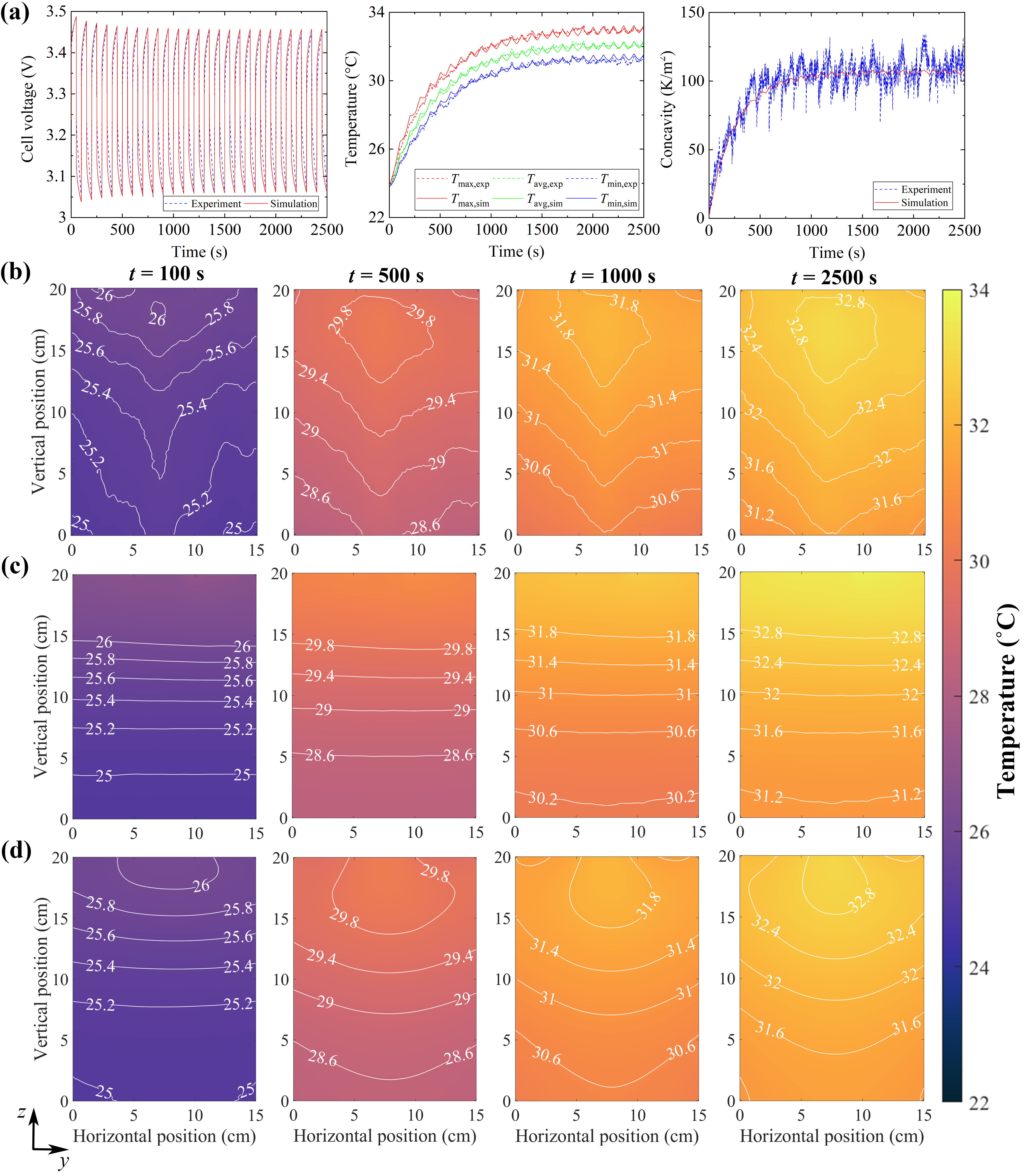}
\caption{Experimental and simulation results for a 20 Ah LFP pouch cell under 4C--100s square-wave cycling around an initial 30\% state of charge. (a) Cell voltage, temperatures and horizontal temperature concavity. (b) Experimental measurements of surface temperature after 100 s, 500 s, 1000 s, and 2500 s of 4C-100 s cycling. (c) Model training results based on the experimental data shown in panel (a) and the streamlined model by Chu \emph{et al.}\ \cite{Howie:2020pa}. (d) Model training results based on the experimental data from (a) and the model proposed in this work. The battery tabs (not shown) are on the top edge of the thermal images in (b)-(d). }
\label{fig:exp_periodic}
\end{figure}

Typical experimental results for a cell cycled from an initial 30\% SOC under a square-wave current having 4C amplitude and \SI{100}{s} period---abbreviated henceforth as 4C--100s@30\%---are shown in Fig.~\ref{fig:exp_periodic} and Supplemental Video 1. 
Before $t=100$ \si{s}, the maximum cell temperature occurs in the areas directly adjacent to the tabs, and a temperature gradient develops primarily in the vertical direction. Near $t=100$~\si{s}, a small `hot spot' --- a point maximum of temperature on the cell surface --- appears near the top edge of the cell and begins to move downward along the vertical centre line. Between $t=100$~\si{s} and $t=500$~\si{s}, a domain of higher temperature around the hot spot gradually expands, eventually achieving a relatively stable shape and location when $t>500$~\si{s}. The average cell temperature also arrives at a state in which it fluctuates around a relatively constant elevated value after about $1500$~\si{s}. The mean surface temperature in this `periodic steady state' is determined by the balance of Joule heating and convection from the cell surface, and the temperature fluctuations arise from entropic effects associated with the cell reaction. In the periodic steady state, the hot spot generally swells and contracts when applied currents have opposite signs; the hot spot expands when the reaction entropy effect is exergonic, and contracts when it is endergonic. 

The strong and relatively isolated correlation between solid-phase diffusion and horizontal temperature concavity (\emph{cf}. Fig.\ \ref{Fig.2}) justifies the estimation of solid-state diffusivity with temperature measurements. Taking the concavity of the temperature distributions into account, model parameters were estimated by fitting 4C--100s@30\%, 4C--100s@50\%, and 4C--100s@70\% data using maximum likelihood estimation, as detailed in the Methods section. The parameters resulting from this estimation process are listed in Table~\ref{tab:fitted_values}, and the fits for 4C--100s@30\% are also plotted on Fig.\ \ref{fig:exp_periodic}(a). Corresponding to the experimental thermograms, Figs. \ref{fig:exp_periodic}(c) and (d) show simulation results for 4C--100s@30\% using the prior streamlined model \cite{Howie:2020pa} and the proposed model, respectively, both based on the best-fit properties provided in Table~\ref{tab:fitted_values}. 

It is clear that the model incorporating solid-state diffusion captures more features of the surface-temperature distribution. This is because solid-state diffusion introduces an additional time constant to the system. Two key relaxation times are visible in Fig.\ \ref{fig:exp_periodic}(a) besides $t_\textrm{d}$: a reaction relaxation time $t_\textrm{rxn}$,
 and a thermal time constant $t_\textrm{th}$.
 A dimensional analysis shows that
\begin{equation}
    t_\textrm{rxn} \sim  \frac{2RT \bar{Q} }{F k_U a i_0^\theta \delta} 
    \approx 40~\si{s}~~~\textrm{and}~~~t_\textrm{th} \sim \frac{\tilde{C}_p \delta}{h} \approx 1400~\si{s},
\end{equation}
where $R$ is the gas constant, $T$, the ambient temperature, and $F$, Faraday's constant; $\bar{Q}$ is the rated capacity of the cell per unit of superficial electrode area ($20~\si{Ah}/300~\si{cm^2}$) and $\delta$, the cell thickness (\SI{7}{mm}); the remaining parameters are defined in Table \ref{tab:fitted_values}. 
The short-time relaxation of cell voltage is controlled by $t_{\textrm{rxn}}$. This is an electrical time constant, which arises from the fact that the slope of the cell's open circuit potential (OCP) with respect to its SOC acts like an admittance, and the exchange-current density acts like a conductance. Thus the interface relaxes somewhat like a parallel RC circuit, with the OCP slope providing the (inverse) capacitance, and interfacial charge-transfer kinetics providing the resistance. Notably, this time constant has not been considered in most asymptotic analyses of the Doyle--Fuller--Newman model \cite{Moura:2017Bat,Sulzer:2019Fast}. The thermal relaxation time $t_\textrm{th}$ is much longer than the other time constants. It controls the relaxation of the voltage envelope and the average cell temperature. As mentioned earlier, the solid-state diffusion time constant $t_\textrm{d}$ controls how the horizontal temperature concavity relaxes.

Parameter values in Table \ref{tab:fitted_values} are similar to those found by fitting with the streamlined model \cite{Howie:2020pa}, with three notable exceptions---ionic conductivity, reaction current density, and thermal conductivity. Since work by Chu \emph{et al.}\ \cite{Howie:2020pa} ignored the concentration polarization caused by solid diffusion, the resulting voltage drop could only be attributed to poor effective ionic conductivity in the electrolyte, which had to be underestimated by \emph{ca}. 50\% to fit the cell-voltage response. Reduced ionic conductivity increases the amount of Joule heating, causing an overestimation of exchange current density in order to lower interfacial resistance and match the temperature. The change in thermal conductivity owes in part to the inclusion of hot-spot position in the cost function used during parameter optimization, as discussed in the Methods section. Separate simulations showed that changing the electrode's effective thermal conductivity tunes the vertical position of the hot spot on the cell surface. This change led to fitted thermal conductivities of the order of \SI{1}{W.m^{-1}.K^{-1}}, placing results in good agreement with independent thermal-characterisation tests undertaken on similar electrode materials \cite{Zhang:2014Si,Werner:2017Th}. 

Among all the battery properties, one expects on a theoretical basis that the volumetric exchange-current density $ai_0^\theta$, OCP gradient $k_U$, diffusion time $t_\text{d}$, and entropy change $\Delta S$ may vary with cell SOC, while the other material properties should be nearly independent of it. In the range of SOC studied here, however, the variations of fitted $ai_0^\theta$ and $t_\text{d}$ with SOC were also minimal. The apparent constancy of $t_\text{d}$ in this range is qualitatively confirmed by the horizontal temperature-profile concavity data shown in Figure S5, which are nearly invariant with respect to cell SOC at a given current density. For A123 20 Ah LFP pouch cells, it appears that a single, SOC-independent diffusion coefficient and exchange-current density suffice to describe measured voltage and temperature behaviour up to 4C.

For validation purposes, simulations using the fitted parameters were also compared with experimental results at conditions not used for fitting, with different C-rates and cycling periods, specifically 4C--50s and 2C--100s cycling, at each SOC. Figure\ S5 provides the experimental and simulated voltage and temperature responses for two validation tests at 30\% SOC (2C--100s@30\% and 4C--50s@30\%), and two 4C--100s parameter-estimation tests at 50\% and 70\% SOC. Measured and predicted thermal images under various square-wave cycling profiles at $t=2500~\si{s}$ are plotted in Figure S6. The root-mean-square errors comparing simulations with experiments are \SI{0.2}{K} for temperature and \SI{5.0}{mV} for voltage.

\section{Cross-scale effects of non-uniform temperature}

The model parameterized above can, with minimal modification, be expanded to full discharge simulations that retain most of parameters estimated from the square-wave-excitation cycling tests. As mentioned before, most parameters vary negligibly with SOC. Full discharges were simulated by leaving every parameter constant apart from the local OCP gradient and the entropy change, which were replaced by local functions of SOC gathered from either full-cell measurements (OCP) or manufacturer-supplied data (entropy change). A description of the model parameterization is available in Supplemental Note 4; a complete set of parameters is given in Table S2; details of how OCP and entropy were handled are discussed in the Methods section. 

\begin{figure}
\centering  
\includegraphics[width=14cm]{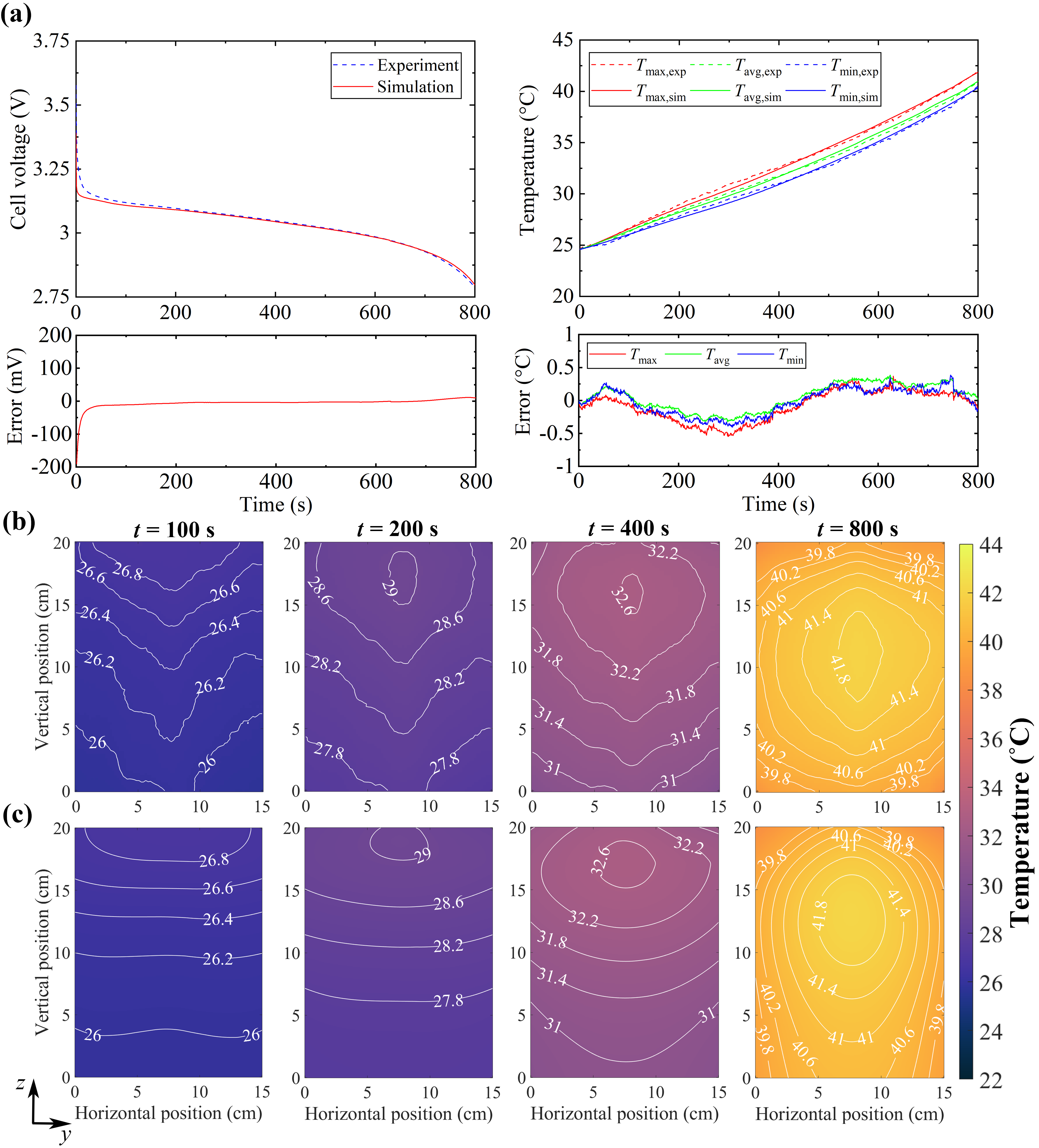}
\caption{Model parameters extracted from square-wave-current perturbations of a 20 Ah LFP pouch cell extrapolate to simulate complete constant-current discharge accurately. (a) Cell voltage and maximum, surface-average, and minimum temperatures during full discharge at 80 A (a 4C rate). (b) Experimental thermograms showing the surface-temperature distribution at various times. (c) Simulation results. The battery tabs (not shown) are on the top edge of the thermal images in panels (b) and (c).}
\label{fig:fulldischarge}
\end{figure}

The maximum C-rate explored in this work is 4C, a fairly challenging test of the accuracy and generality of our approach. The LFP pouch cell was discharged at constant current from 100\%~SOC to 0\%~SOC, after first charging the cell to \SI{3.6}{V} with a `CC-CV' protocol (in which a 4C current was applied until the voltage reached \SI{3.6}{V}, at which it was held until current decayed to C/100) and then resting for an additional hour. The cell voltage, current, and surface-temperature distribution were measured at \SI{1.0}{Hz}. Figure\ \ref{fig:fulldischarge}(a) and (b) show the battery voltage and temperature response during the 4C discharge. 
Initially ($t<100$~\si{s}), similarly to the square wave cycling tests, the region close to the battery's tabs rapidly warms up due to the higher current density there, forming a hot spot near the cell's top edge. Between $t=100$~\si{s} and $t=200$~\si{s}, the size of the hot spot gradually expands vertically; the point of maximum temperature detaches from the top edge at $t=200$~\si{s}. As the discharge continues, the hot spot moves down towards the centre of the cell and grows in size. The centre of the hot spot shifts downward substantially, passing through \emph{ca.}\ $z=160$~\si{mm} at $t=400$ s and $z=110$~\si{mm} at $t=800$~\si{s}. 

The average surface temperature increases by \SI{17}{\degreeCelsius} during the 4C discharge. At the end of discharge, the surface-temperature distribution spans \SI{1.5}{\degreeCelsius} between the maximum temperature, at the hot spot, and the minimum temperature, at the bottom edge. Simulated transient surface-temperature fields are shown in Figure\ \ref{fig:fulldischarge}(c) and a real-time comparison between the thermography test and model simulation is provided in Supplemental Video 2; the cell temperature rise and distribution are well matched between the model and experiment throughout the discharge. 

\begin{figure}
\centering  
\includegraphics[width=13cm]{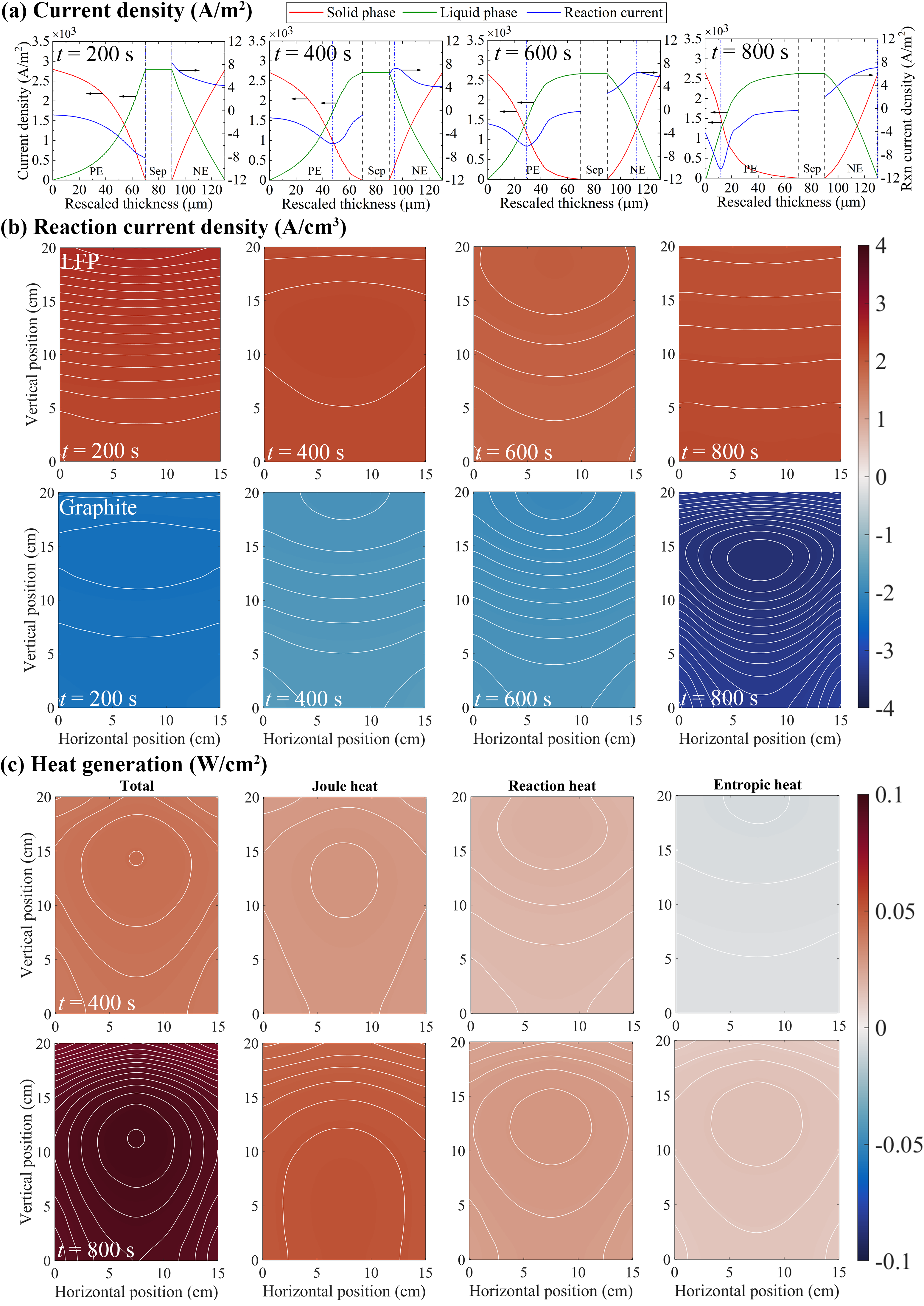}
\caption{Fine-grained simulation results for 4C constant-current discharge. (a) Spatiotemporal variations of solid-phase, liquid-phase, and reaction current along an axis through the hot spot at horizontal position $80~\si{mm}$ and vertical position $140~\si{mm}$. Labels PE, Sep and NE indicate positive-electrode, separator, and negative-electrode domains, respectively. (b) Local reaction current density in central cross-sections of the LFP and graphite electrodes through the blue dash-dotted lines in panel (a). The contour lines are spaced at intervals of \SI{20}{mA.cm^{-3}}. (c) In-plane total, Joule, reaction, and entropic heat generation in the LFP and graphite electrodes at $t=400~\si{s}$ and $t=800~\si{s}$. Each location in these surface plots shows the integral of local heat generation across the entire electrode thickness; the contour lines are spaced at intervals of $1$ \si{mW.cm^{-2}}.}
\label{fig:sim_results}
\end{figure}

Figure\ \ref{fig:sim_results}(a) shows spatial and temporal variations of solid-phase, liquid-phase, and reaction current density through-plane (\emph{i.e.}\ normal to the $y$-$z$ plane shown in Fig. \ref{fig:fulldischarge}) through the in-plane ($y$-$z$) location of the hot spot.
Generally, the magnitude of liquid-phase current density increases toward the separator, while the solid-phase current drops. Reaction current is associated with the concavity of these profiles. Extrema of the reaction current are observed in both electrodes at all times. Electrochemical reactions in the electrodes are initially favoured at the electrode/separator interfaces, so the extrema originate near the separator in both electrodes. As discharge progresses, the extrema become peaks, and the reaction front moves over time toward the current collectors. The presence of peaks in reaction current owes to the solid-phase diffusion limitation, a phenomenon familiar from one-dimensional DFN models in the literature \cite{Fuller:1994Si}. Diffusion also controls the rate at which the reaction-current peaks move towards the current collectors. Because the graphite electrode is thinner, the extremum of reaction current has reached the current collector at \SI{800}{s}, whereas the LFP reaction-current distribution still exhibits a peak. But because the LFP electrode is thicker, the peak in reaction current is narrower, and has a higher magnitude; the separator side of the LFP is fully discharged just after \SI{400}{s}, and can no longer sustain reaction current.  

The current densities also vary substantially in-plane direction. Figure \ref{fig:sim_results}(b) shows in-plane reaction-current distributions along cross-sections cut through the instantaneous extrema of reaction current (these positions are indicated with blue dash-dotted lines in Figure \ref{fig:sim_results}(a)). 
Peak-to-valley variation relative to the average ranges from $ca.$ 2\% to 20\%, and differs between the two electrodes at a given instant, despite the solid-phase diffusion times being the same. 
When the extremum in reaction current resides at the separator or current collector, the reaction current has relatively uniform gradient top-to-bottom. Otherwise, the $x$-location of the peak, \emph{cf.} Fig. \ref{fig:sim_results}(a), is also associated with a peak in the $y$-$z$ plane. Again, the peak is sharper in the thicker electrode. 

Figure\ \ref{fig:sim_results}(c) presents the in-plane distributions of instantaneous heat generation, integrated across the thickness of the whole cell after \SI{400}{s} and \SI{800}{s} of 4C discharge. The total heat generated is broken down into contributions from Joule heat (arising from bulk resistances in Eqn.\ (\ref{eq:energy_bal}) of the Methods section), reaction heat (interfacial resistances), and entropic heat. The local maxima in total through-plane heat generation are similar to the hot-spot locations in Fig.\ \ref{fig:fulldischarge}. The total heat generation generally increases as the cell is discharged. Joule heating accounts for about half of the total: at \SI{400}{s} it is relatively uniformly distributed in-plane; at \SI{800}{s} its distribution is dominated by an increase from the top of the cell to the bottom. Reaction heat and entropic heat, on the other hand, both come to relatively sharp peaks in the $y$-$z$ plane. They interfere destructively at \SI{400}{s}, where entropic heat is negative, and constructively at \SI{800}{s}. Nevertheless the sum of the two is always positive; reaction heat and entropy together dominate the placement of the peak in total heat generation. Although the reaction-current distributions in Fig.\ \ref{fig:sim_results}(b) are difficult to parse, it does appear that the average position of the extreme reaction currents correlates with the maximum in total heat generation. Thus it can be concluded that the position of the hot spot is largely controlled by nonuniform in-plane reaction current.

\section{Conclusions}
\label{S:5}

Multiple processes within a battery have distinctive heat signatures, including electrochemical reactions, interfacial kinetics, and Joule heating. Thus, in large lithium-ion pouch cells, surface temperature can be used as an effective probe to provide microscopic understanding. Using a combination of lock-in thermography and physics-based modelling, we characterised several material properties by inverse modelling of experimental tests with square-wave applied currents. Significant in-plane temperature non-uniformity was observed, and was attributed to a balance of the heterogeneous distributions of local charge state, interfacial Joule heating, and ohmic heat generation. Importantly, solid-phase diffusion of intercalated lithium was found to have a distinctive macroscopic heat signature, leading to concavity in the temperature distribution along an axis parallel to the current collectors but perpendicular to the battery's tabs.

We showed that many inferred properties could be assumed constant over a very wide range of states of charge. A model that kept most parameters fixed but included a lookup table of pseudo-OCP data accurately simulated the cell voltage and thermal response during galvanostatic full discharges at 4C. Joule heating and reaction heat are generally comparable in magnitude; the position of the maximum temperature on the cell surface is controlled primarily by the microscopic reaction distribution and solid-phase diffusion. Since large-format lithium-ion cells are favoured for high-energy-density packs, detailed knowledge about non-uniform thermal states is critical to the understanding of battery performance and cycle life, particularly during fast charging or discharging. 

\section*{Methods}
\label{S:2}

\subsection*{Electrochemical testing}
All tests reported were performed using commercial \SI{20}{Ah} LFP pouch cells (AMP20M1HD-A, A123 Systems). The equilibrium open-circuit potential (OCP) of the cells was measured with a pseudo-OCP approach at C/25 applied current between \SI{3.6}{V} (identified as 100\% SOC) and \SI{2.5}{V} (0\% SOC). The charge capacity was determined via Coulomb counting under a CC-CV protocol at 1C until the current decayed to C/100. When a cell was set to a certain initial SOC, it was first charged to 100\% SOC with a CC-CV protocol, then discharged to the required setpoint by Coulomb counting at 1C. All these tests were conducted using an automated battery test system (Series 4000, Maccor Inc.). In the lock-in thermography tests, the cells placed in the test rig were charged and discharged using a high-power bipolar power supply (BOP 10-100MG, KEPCO Inc.). Square-wave-excitation cycling data were gathered based on initial SOCs of 30\%, 50\% or 70\%, with applied current at 2C or 4C and periods of $50~\textrm{s}$ or $100~\textrm{s}$ \cite{Howie:2020pa}. Full-discharge experiments were conducted from 100\% SOC to 0\% SOC with applied current at 0.2C, 1C, 2C and 4C.

\subsection*{Pouch cell disassembly}
To determine the correct physical dimensions and layer structure to be used in finite-element modelling of the LFP pouch cell, a cell was disassembled in a glove box after being fully discharged to \SI{2.5}{V}. Photographs of single layers of the cathode and anode extracted from the pouch cell are shown in Figure S1. The geometric parameters of the battery and its components were measured and are summarized in Table S1. 

\subsection*{Lock-in thermography and thermal image analysis}
Lock-in thermography was conducted following the method of Chu \emph{et al.} \cite{Howie:2020pa} using a thermal imaging camera (A35sc, FLIR Systems). Spatiotemporal temperature data was gathered by image processing, including reference-temperature data averaged over a piece of black felt, labeled as the `ambient' rectangle in Fig.\ \ref{fig:experimental_rig}. The ambient-spot temperature from thermography was calibrated against measurements from a thermocouple ($\pm$\SI{0.1}{\degreeCelsius} accuracy, Type~T, Omega Engineering) placed behind a piece of felt at that location, also labelled in the Fig.\ \ref{fig:experimental_rig} inset. The resulting baseline temperature was subtracted from each pixel of the image during data processing.  

\subsection*{Battery model}
\label{S:3.1}
Transient surface-temperature and voltage profiles were simulated using a reduced-order continuum model, which was derived from an extended Doyle--Fuller--Newman model that incorporates a local energy balance, as detailed in Supplemental Note 2. This 3D model was applied throughout the pouch interior. As described in section \ref{results:nonuniform} above, a simplified model, which neglects diffusion in the electrolyte but retains solid-state diffusion in the electrodes, was deemed sufficient to fit the data. 

Within the macroscopic volume spanned by a given electrode, this reduced-order model considers ohmic charge balances in the solid and liquid phases, respectively, such that
\begin{equation}
\label{eq:charge_solid}
\nabla\cdot\vec{i}_\textrm{s}=-ai,
\end{equation}
\begin{equation}
\label{eq:charge_liquid}
\nabla\cdot\vec{i}_\textrm{l}=ai,
\end{equation}
\begin{equation}
\vec{i}_\textrm{s}=-\sigma \nabla \phi_\textrm{s},
\end{equation}
\begin{equation}
\vec{i}_\textrm{l}=-\kappa \nabla \phi_\textrm{l}, 
\end{equation}
in which $\vec{i}_k$ and $\phi_k$ are respectively the current density and electrical potential in phase $k$ (with subscript `l' designating liquid and `s', solid), $a$ is the pore surface area per unit electrode volume, and $i$ is the current density across the pore surface, defined such that anodic currents are positive. The effective electronic conductivity of the solid is $\sigma$, and the effective ionic conductivity of the liquid, $\kappa$. Ionic conductivity was taken to vary with absolute temperature $T$ according to
\begin{equation}
\kappa=\kappa^\theta +\alpha \left(T-T^\theta \right),
\end{equation}
where $\kappa^\theta$ is the conductivity at reference temperature $T^\theta$ and $\alpha$ expresses its linear variation.

Temperature is distributed across the electrode domains according to a macroscopic thermal energy balance, derived under the assumption that the interpenetrating liquid and solid phases that make up the electrode sit at equal temperatures:
\begin{equation}
\label{eq:energy_bal}
\tilde{C}_p\frac{\partial T}{\partial t} = \nabla\cdot \left(k \nabla T \right)+\sigma \nabla\phi_{\textrm{s}}\cdot\nabla\phi_{\textrm{s}}+\kappa \nabla\phi_{\textrm{l}}\cdot\nabla\phi_{\textrm{l}}+ai\eta+aiT\Delta S.
\end{equation}

\noindent Here $\tilde{C}_p$ is the effective local volumetric heat capacity, $k$ is the effective thermal conductivity, and $\Delta S$ is the reaction entropy of the electrode half-reaction. 

The applied current ($i_{\text{app}}=\vec{n}\cdot\vec{i_s}$) and the electric ground ($\phi_s=0$) are defined at the positive and negative terminals (copper bars, \emph{cf.} Fig.\ S9), respectively. The component of $\vec{i}_\textrm{l}$ normal to interfaces between the current collector and the electrodes was taken to vanish; similarly, the components of $\vec{i}_\textrm{s}$ normal to interfaces between the anode and cathode were taken to vanish. The outer edges of the pouch were taken to be electrically insulating. To bound the thermal portion of the problem, Newton's law of cooling was adopted at the exterior surfaces of the cell: 
\begin{equation}
- \left. \left( \vec{n} \cdot k \nabla T \right) \right|_{t,\textrm{pouch}} =h \left( \left. T \right|_{t,\textrm{pouch}} -T_0 \right),
\end{equation}
in which $h$ is the heat transfer coefficient and $T_0$ is the ambient temperature.

Active particles within each electrode are taken to exist across a microscopic radial dimension $r$ at each point within the electrode. Within these spherical particles of radius $r_0$, the concentration $c_\textrm{s}$ of intercalated lithium is taken to satisfy Fick's law
\begin{equation}
\frac{\partial c_\textrm{s}}{\partial t}=\frac{D_{\textrm{s}} }{r^2}\frac{\partial}{\partial r} \left( r^2 \frac{\partial c_\textrm{s}}{\partial r} \right),
\end{equation}
in which $D_{\textrm{s}}$ is the solid-phase diffusivity. This microscopic mass balance is coupled to the macroscopic problem through a boundary condition 
\begin{equation}
-D_\textrm{s} \left. \frac{\partial c_\textrm{s}}{\partial r} \right|_{t,r_0} =\frac{i}{F},
\end{equation}
where $F$ stands for Faraday's constant, and $c_\textrm{s} \left(t,0\right)$ is required to be finite. Thus the local macrosopic interfacial current density $i$ completely specifies the diffusion dynamics within the solid particles.

As justified in Supplemental Note 2, the interfacial reaction currents were taken here to follow linear kinetics,
\begin{equation}
i=i_0\frac{F \eta}{RT},
\end{equation}
where \(i_0\) is the exchange current density and $R$, the gas constant. The temperature dependence of exchange-current density was modeled as \cite{guo2010single}

\begin{equation}
i_{0}=i_0^\theta \exp\left[-\frac{E^\theta}{R}\left(\frac{1}{T}-\frac{1}{T^\theta} \right)\right],\;
\end{equation}
in which $i_0^\theta$ is the exchange-current density at $T^\theta$ and $E^\theta$ is an Arrhenius parameter. Charge transfer is driven by the surface overpotential $\eta$ between the liquid and solid phases, which breaks down as

\begin{equation}
\eta=\phi_\textrm{s}-\phi_\textrm{l}-U,
\end{equation}
in which $U$ is the electrode's equilibrium OCP relative to a reference electrode of a given kind. 

During square-wave cycling tests, the perturbation in SOC is small, allowing it to be  linearized following Chu \emph{et al.}\ \cite{Howie:2020pa}. Whereas the equation of Chu is based on the average SOC within the particle, the diffusion limitation requires that this be replaced by the effective SOC at the particle surface, $q$, defined as
\begin{equation}
    q = \frac{\left. c_\textrm{s} \right|_{t,\textrm{surf}}}{c_{\textrm{s,max}}} .
\end{equation}
Here $\left. c_\textrm{s} \right|_{t,\textrm{surf}}$ is the instantaneous surface concentration at $r=r_0$ within the solid and $c_{\textrm{s,max}}$ is the maximum lithium concentration the solid particles can accept. Taking account of the diffusion limitation's effect on the surface concentration of intercalated lithium, one finds that the OCP satisfies
\begin{equation}
\label{eq:OCP}
U=U_0+k_{U} \left( q  - q_0 \right) +\frac{\Delta S_0}{F}\left(T-T_0\right)+V_{\textrm{hys}}\cdot \text{sgn}(i),
\end{equation}
in which $q_0$ is the average fractional state of charge of the whole pouch cell. The constant parameter $V_{\textrm{hys}}$ is included to describe possible OCP hysteresis during slow charge or discharge \cite{srinivasan2006existence}. Here, a nonzero value of $V_\textrm{hys}$ was included to model LFP cathodes; $V_\textrm{hys}$ was taken to be zero for graphite anodes. Here, $U_0$ is the OCP at $q_0$, $k_{U}$ represents the OCP gradient with respect to fractional SOC at $q_0$, and $\Delta S_0$ is the reaction entropy at $q_0$ and $T_0$, which appears because OCP satisfies the Maxwell relation
\begin{equation}
\Delta S \left( q \right) = F \left(\frac{\partial U}{\partial T} \right)_q.
\end{equation}
Generally we assume that the reaction entropy depends weakly on temperature, and is therefore a function of SOC only. 

The same numerical model was used to simulate full discharges, except the linearized OCP curves from equation \ref{eq:OCP} were replaced with full-cell experimental pseudo-OCP discharge data in the nonlinear form
\begin{equation}
    U  = U_0 \left( q , T_0 \right) + \frac{\Delta S \left( q \right) }{F}  \left( T - T_0 \right) ,
\end{equation}
where the functions $U_0$ and $\Delta S$ come from experiments. (Note that no hysteresis term is present here because only discharges were modelled.) These data were used assuming the anode as a reference potential: thus the model used $U$ as the equilibrium voltage in LFP, and assumed the graphite OCP to be ground (0 V). In the model, reversible heating was computed from full-cell $\Delta S$ data by equally apportioning the reaction entropy between the two electrodes. This could be refined if reference-electrode measurements were available, but practically, the thinness of the cell normal to the electrodes means that an unequal distribution of reversible heat is difficult to discern. The full-cell OCP and $\Delta S$ data are provided in Figures S7 and S8, respectively. A detailed description of the model parameterization is available in Supplemental Note 4, and a complete set of model parameters is given in Table S2. 

When solving the model, computational speed can be improved by a scaling analysis of the governing equations. By applying the scaling argument put forward by Chu \emph{et al.} \cite{Howie:2020pa}, the multi-layer internal geometry was homogenised, allowing the electrochemical model to be solved across a single representative layer. The scaling procedure and details of the simulated geometry are described in Supplemental Note 3. 

\subsection*{Parameter estimation}
\label{S:3.3}

Inverse modeling was based on iterative solution of the transient battery model. The parameter vector 
\begin{equation}
X=\left[ai_0^\theta,\;E_{i0},\;\kappa,\;\alpha,\;k_U,\;\frac{r_0^2}{D_\textrm{s}},\;\Delta S,\;\frac{h}{\tilde{C_p}},\;k\right]
\end{equation}
was identified using a nonlinear least-squares fitting algorithm that minimised the error between measurements and simulations (respectively denoted with superscripts `exp' and `sim') at each time-step $i$ of voltage, $V_i$; maximum, minimum, and surface-averaged temperatures $T_{\textrm{max},i}$, $T_{\textrm{min},i}$, and $T_{\textrm{avg},i}$, respectively; concavity of the temperature distribution $k_{\textrm{c},i}$; and the position of the hot spot in the $yz$ plane, $\left( y_\textrm{hot} , z_\textrm{hot} \right)$. The objective function $f$ for the minimisation was expressed as a sum over all $N$ entries in the time series, as
\begin{equation}
\begin{aligned}
f = \sum\limits_{i = 1}^N &{\left[ {{{\left( {\frac{{V_i^{\text{sim}} - V_i^{\text{exp}}}}{{\Delta V^{\text{exp}}}}} \right)}^2} + {{\left( {\frac{{T_{\text{max},i}^{\text{sim}} - T_{\text{max},i}^{\text{exp}}}}{{\Delta T_{\text{max}}^{\text{exp}}}}} \right)}^2} + {{\left( {\frac{{T_{\text{min},i}^{\text{sim}} - T_{\text{min},i}^{\text{exp}}}}{{\Delta T_{\text{min}}^{\text{exp}}}}} \right)}^2}} \right.}\\
&\left. { + {{\left( {\frac{{T_{\text{avg},i}^{\text{sim}} - T_{\text{avg},i}^{\text{exp}}}}{\Delta {T_{\text{avg}}^{\text{exp}}}}} \right)}^2} + {{\left( {\frac{{k_{c,i}^{\text{sim}} - k_{c,i}^{\text{exp}}}}{{\Delta k_{c}^{\text{exp}}}}} \right)}^2}} + \frac{{{{(y_{\text{hot},i}^{\text{sim}} - y_{\text{hot},i}^{\text{exp}})}^2} + {{(z_{\text{hot},i}^{\text{sim}} - z_{\text{hot},i}^{\text{exp}})}^2}}}{{ 
L_y L_z}} \right] ,
\end{aligned}
\end{equation}
where $\Delta V^\textrm{exp}$, $\Delta T_\textrm{max}^\textrm{exp}$, $\Delta T_\textrm{min}^\textrm{exp}$, $\Delta T_\textrm{avg}^\textrm{exp}$ and $\Delta k_c^\textrm{exp}$ denote the ranges of $V$, $T_\textrm{max}$, $T_\textrm{min}$, $T_\textrm{avg}$ and $k_c$ in the experiment. The width and length of the pouch cell are $L_y$ and $L_z$.  The temperature concavity was calculated at the hot spot location by fitting the horizontal temperature profile through the hot spot (see Figure S3) with a quadratic polynomial, as described in Supplemental Note 1. 

\section*{Acknowledgements}
The authors gratefully acknowledge funding from the EPSRC Translational Energy Storage Diagnostics (TrEnDs) project (EP/R020973/1), the Faraday Institution Multiscale Modelling project (subaward FIRG003 under grant EP/P003532/1) and the STFC Futures Early Career Award. The authors also thank Andrew Wang for his help with pouch-cell disassembly. 





\bibliographystyle{elsarticle-num-names}
\biboptions{sort&compress}
\bibliography{bibliography.bib}


\section*{Supplementary material}

# Supplementary Information

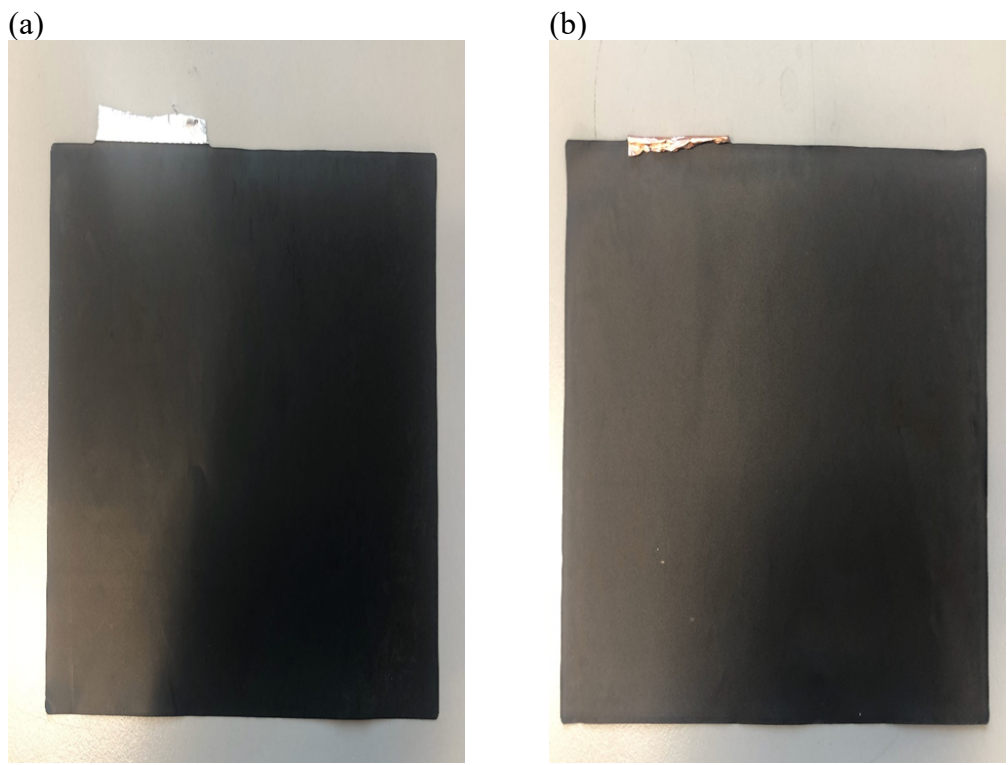

**Figure S1.** A single layer of electrodes in the LFP pouch cell. (a) Cathode; (b) Anode.

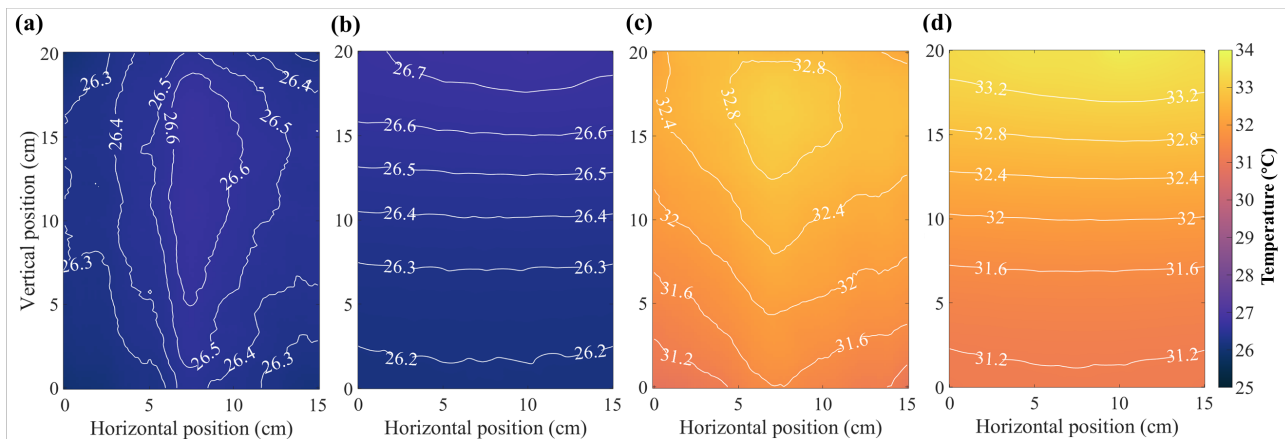

**Figure S2.** Thermal images of the LFP pouch cell at $t$=2500 s under different test conditions. (a) Square-wave cycling with 2C-100 s applied current at 30% SOC. (b) Simulation of (a) with the streamlined model.[1] (c) Square-wave cycling with 4C-100 s applied current at 30% SOC. (d) Simulation of (c) with the streamlined model.[1] The battery tabs (not shown) are on the top edge of the thermal images.

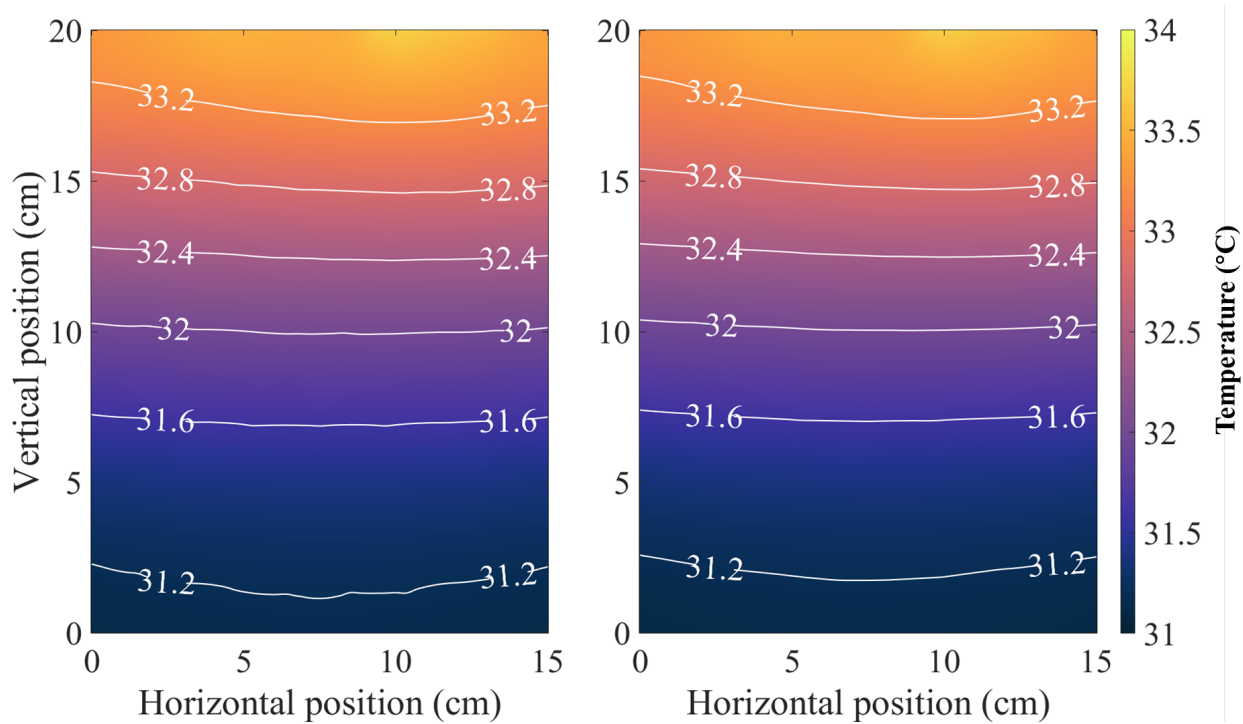

**Figure S3.** Simulation results of LFP pouch cell by the streamlined model[1] with different reaction kinetics equations. (a) Linear reaction kinetics; (b) Original Butler–Volmer reaction kinetics. The simulated thermal images are at $t$=2500 s and 30% SOC with 4C-100 s applied current. The battery tabs (not shown) are on the top edge of the thermal images.

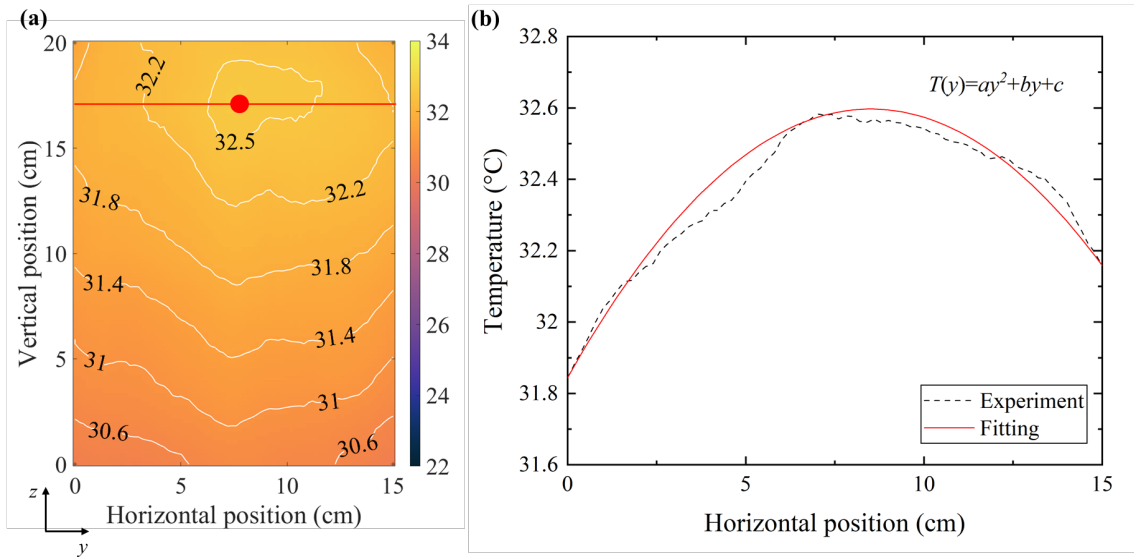

**Figure S4.** Computation of the cell's horizontal temperature concavity from data. (a) Surface-temperature variation yielded by a thermography image at $t$=2500 s from a square-wave cycling test with 4C-50 s applied current at 30% SOC (tabs at the top of the image), showing the location of the hot spot (red •) and an axis passing through it parallel to the $y$ axis (red –). (b) Fit of the temperature variation along the horizontal axis through the hot spot with a quadratic polynomial, which is forced through the temperatures at the left and right edges and the hot spot. The horizontal temperature concavity relates simply to parameter $a$, as described below in Supplemental Note 1.

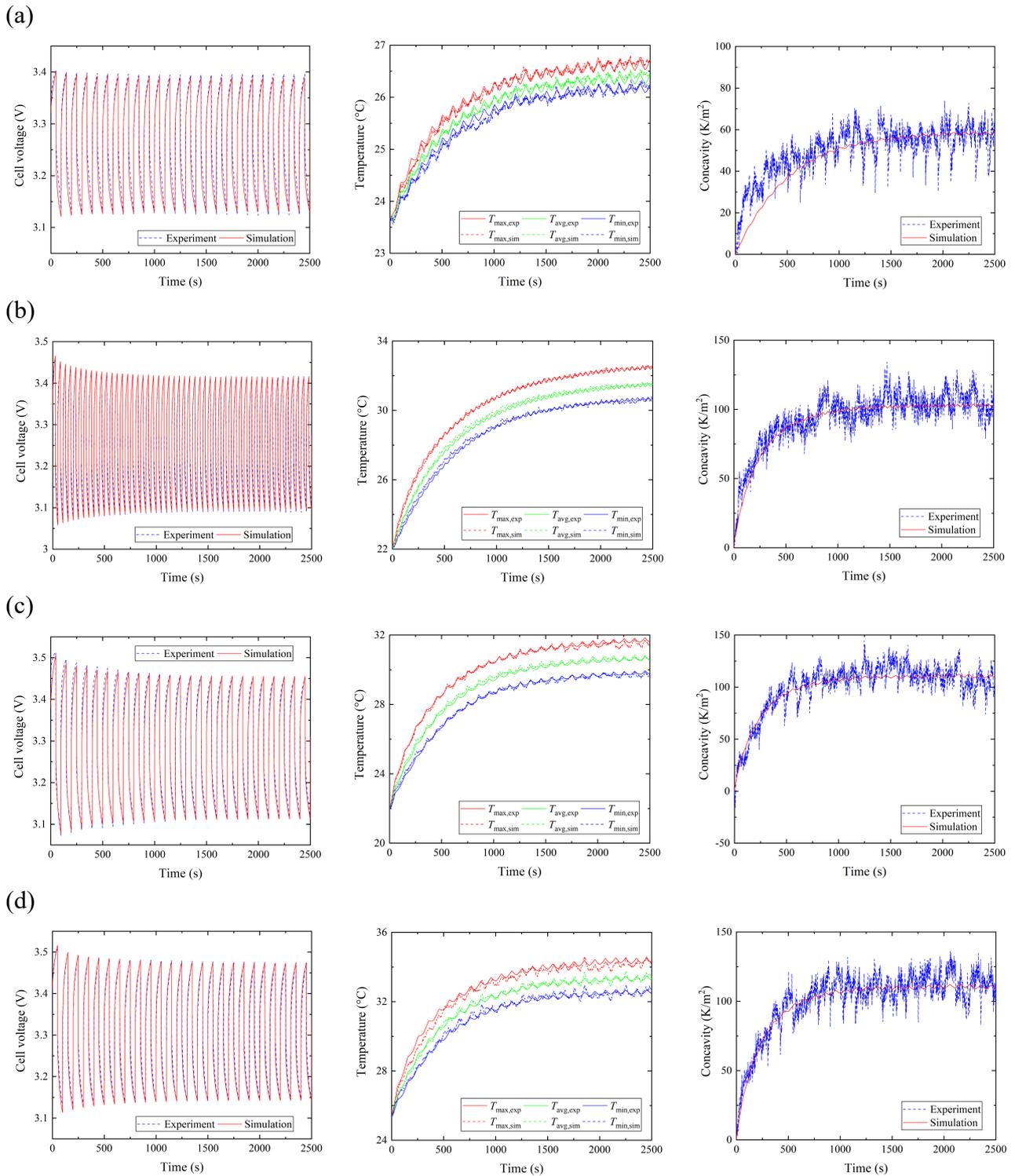

**Figure S5.** (a,b) Validation of parametrization yielded by the fits in Fig. 3, showing cell voltage, surface temperatures, and horizontal temperature concavity measured and predicted under square-wave cycling at (a) 2C-100 s@30% SOC and (b) 4C-50 s@30% SOC cycling. (c,d) Data fits resulting from parametrization of experimental data at (c) 4C-100 s@50% SOC and (d) 4C-100 s@70% SOC.

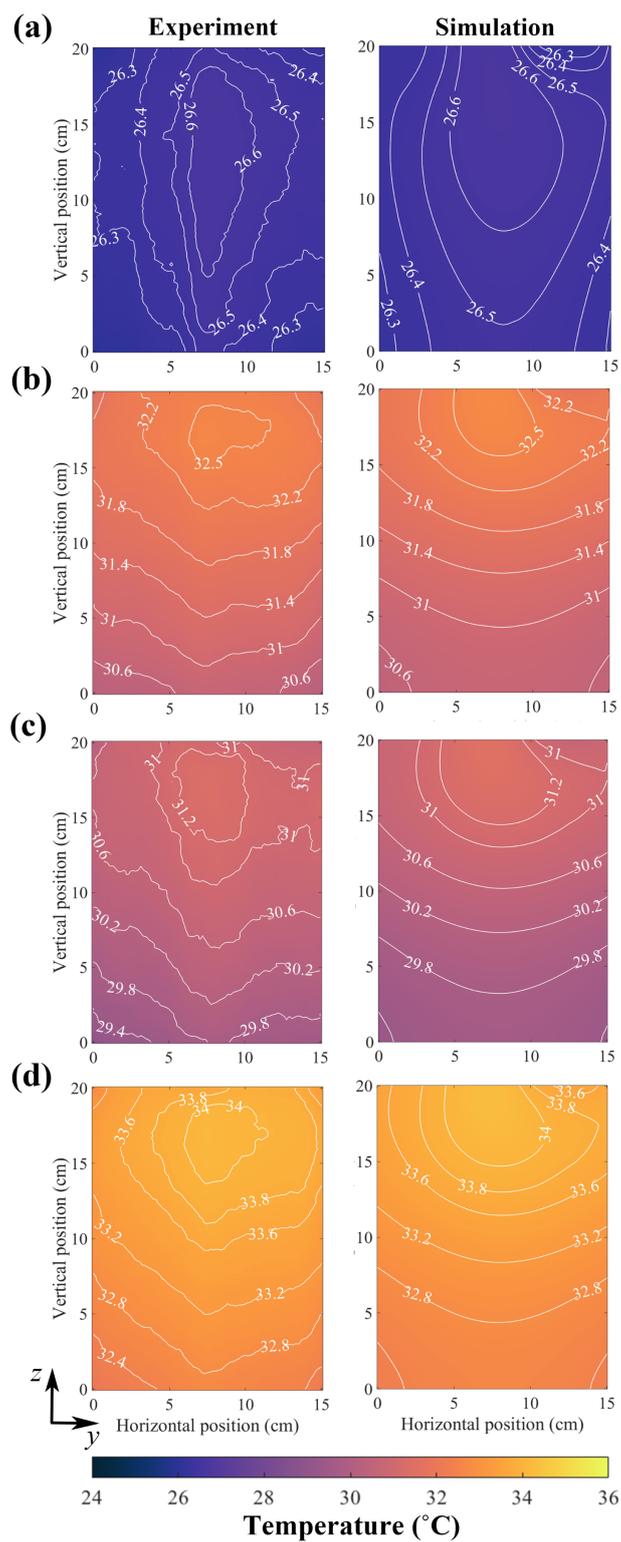

**Figure S6.** Experimental (left) and simulated (right) thermal images ($t$=2500 s) of LFP pouch cells with different applied square-wave currents and initial SOCs. (a) 2C-100 s @30%SOC. (b) 4C-50 s @30%SOC. (c) 4C-100 s @50%SOC. (d) 4C-100 s @70%SOC. The battery tabs (not shown) are on the top edge of the thermal images.

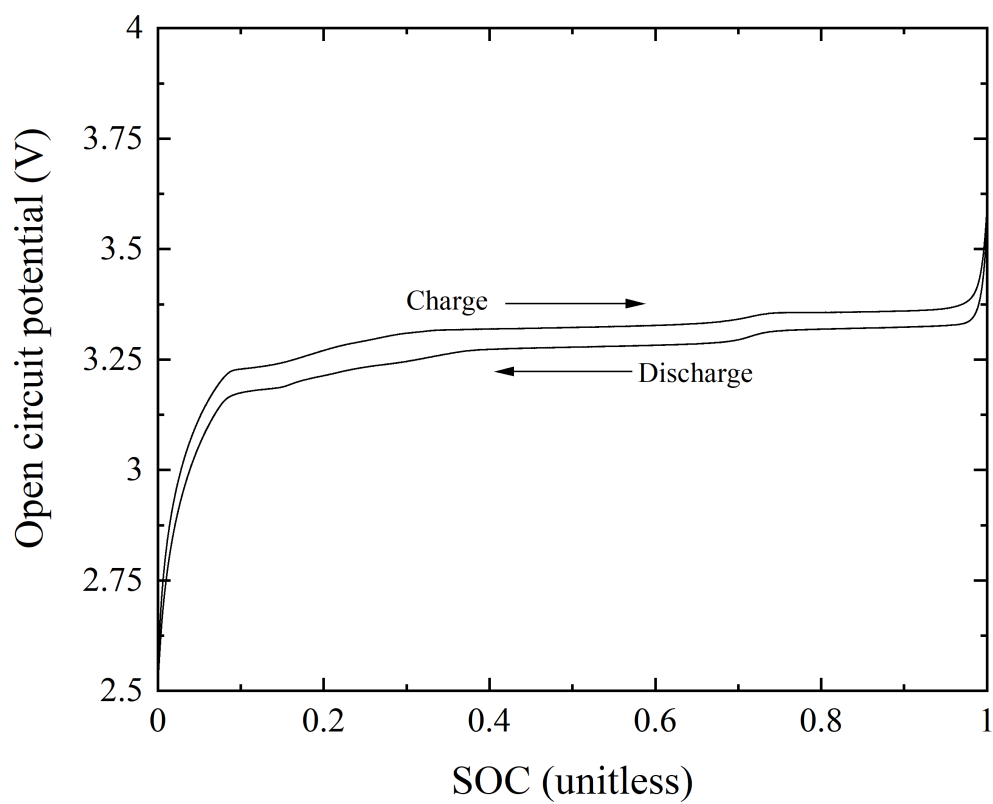

**Figure S7.** Pseudo-open circuit potential (OCP) of the LFP pouch cell, measured at C/25 constant current in a thermal chamber at 25.0 °C.

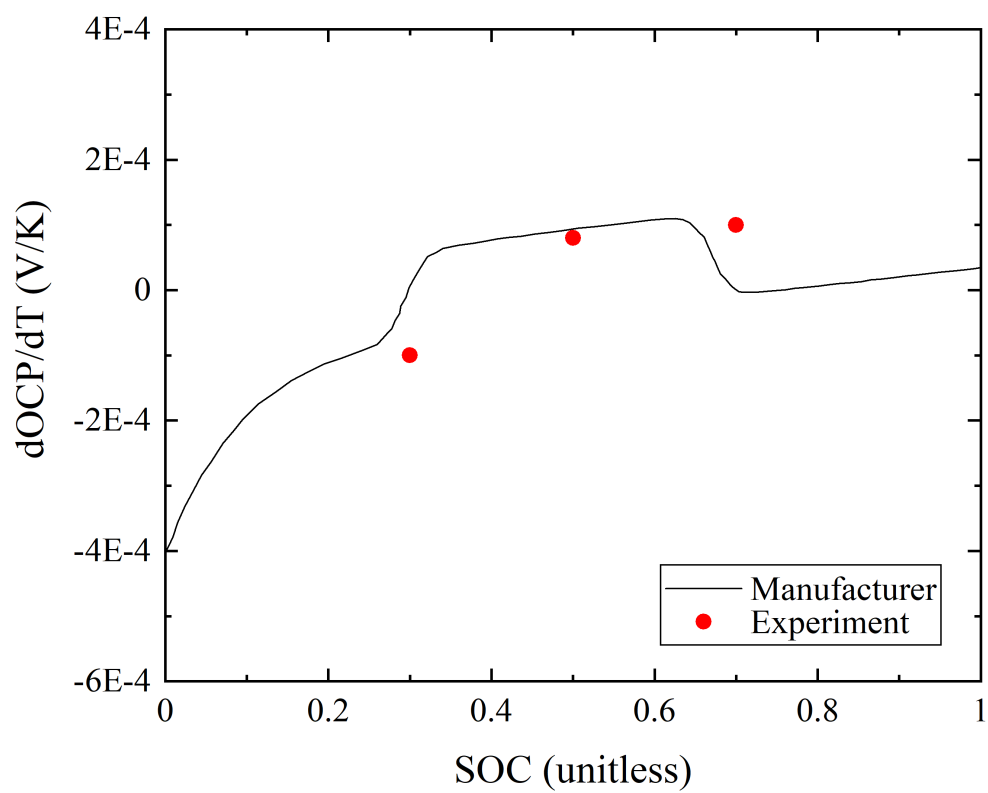

**Figure S8.** Temperature derivative of LFP pouch cell OCP. The experimental data points were obtained from square-wave cycling tests at 30%, 50%, and 70% SOC.

**Table S1** Specifications of the A123 LFP pouch cell.

| Parameter | Symbol | Unit | Value |
|---|---|---|---|
| Battery length | $L$ | mm | 200 |
| Battery width | $W$ | mm | 150 |
| Battery thickness (excluding pouch thickness) | $\delta$ | mm | 6.5 |
| Tab length | $L_{tab}$ | mm | 25 |
| Tab width | $W_{tab}$ | mm | 48 |
| Tab thickness | $\delta_{tab}$ | mm | 0.25 |
| Positive current collector half-thickness | $\delta_{pe,cc}$ | μm | 12.5 |
| Positive electrode thickness | $\delta_{pe}$ | μm | 70 |
| Separator thickness | $\delta_{sep}$ | μm | 20 |
| Negative electrode thickness | $\delta_{ne}$ | μm | 40 |
| Negative current collector half-thickness | $\delta_{ne,cc}$ | μm | 12.5 |
| Number of stacked unit cells | $N$ | | 42 |

**Table S2** Model parameters used for full-discharge simulations. Electrode porosity and particle radius were assumed identical for the positive and negative electrodes; values labelled 'assumed' are based on previous studies on similar cells[2] and are consistent with other reports from the literature.[3-5,7] The effective electrical and ionic conductivity, solid-state diffusivity, reaction current density, heat capacity and thermal conductivity obtained from the square-wave cycling tests were directly applied in the simulations.

| Parameter | Symbol | Cathode | Anode | Method |
|---|---|---|---|---|
| Porosity (unitless) | $\varepsilon_l$ | 0.50 | 0.50 | Assumed |
| Particle radius (μm) | $r_0$ | 5.0 | 5.0 | Assumed |
| Charge capacity (Ah) | $\overline{Q}$ | 20.0 | | Manufacturer |
| Effective electrical conductivity (S/m) | $\sigma$ | 50 | | Fitted |
| Effective ionic conductivity (S/m) | $\kappa$ | 0.046 | | Fitted |
| Temperature coefficient of $\kappa$ (mS/(m·K)) | $\alpha$ | 2.4 | | Fitted |
| Solid-state diffusivity (m²/s) | $D_s$ | $4.5 \times 10^{-14}$ | | Fitted |
| Local reaction current density (A/m²) | $i_0^\theta$ | 6.0 | | Fitted |
| Reaction activation energy of $i_0^\theta$ (kJ/mol) | $E^\theta$ | 29.5 | | Fitted |
| Heat capacity (J/(m³·K)) | $C_p$ | $2.35 \times 10^6$ | | Fitted |
| Effective thermal conductivity (W/(m·K)) | $k$ | 1.1 | | Fitted |
| Heat transfer coefficient (W/(m²·K)) | $h$ | 12.0 | | Fitted |

## Supplemental Note 1. Horizontal temperature concavity

We call the 'horizontal direction' the direction perpendicular to the battery tabs and parallel to the separator (the direction of the $y$-axis in Fig. S4 or Fig. S6). The effective horizontal temperature concavity was approximated by fitting the temperature distribution in the horizontal direction along an axis through the hot spot on the cell surface, that is, the temperature in the $y$-direction passing through the location of the surface-temperature maximum.

Effective concavity was estimated by assuming that the horizontal temperature profile through the hot spot could be approximated by a quadratic function,
$$T(y) = ay^2 + by + c. \tag{S1}$$
The coefficients in equation S1 were determined by forcing this function through three temperatures along the horizontal axis through the hot spot (cf. Fig. S4a): the temperatures at the left and right edges, $T(0) = T_0$ and $T(L_y) = T_L$, respectively, and the temperature at the position of the hot spot, $T(y_{hot}) = T_{hot}$. The results of this fitting process show that the effective horizontal concavity of the temperature profile, $k_c$, is

$$k_c = \tfrac{1}{2}|T''(y_{hot})| = |a| = \frac{1}{y_{hot}(L_y - y_{hot})}\left\{T_{hot} - \left[T_L \frac{y_{hot}}{L_y} + T_0\left(\frac{L_y - y_{hot}}{L_y}\right)\right]\right\}. \tag{S2}$$

## Supplemental Note 2. Battery model

The original Doyle-Fuller-Newman (DFN) model considers charge and mass balances in both the solid and liquid phases within porous electrodes, with the heterogeneous reaction rate within the material calculated via Butler-Volmer kinetics, and the intercalation dynamics governed by spherical diffusion. The governing equations at interior points are written as follows:

Charge balance in solid: $\quad\quad\quad\quad\quad\quad \nabla \cdot \vec{i}_s = -ai \quad\quad\quad\quad\quad\quad$ (S3)

Charge balance in liquid: $\quad\quad\quad\quad\quad\quad \nabla \cdot \vec{i}_l = ai \quad\quad\quad\quad\quad\quad$ (S4)

Mass balance in solid particles: $\quad\quad \dfrac{\partial c_s}{\partial t} = \dfrac{D_s}{r^2}\dfrac{\partial}{\partial r}\left(r^2 \dfrac{\partial c_s}{\partial r}\right) \quad\quad$ (S5)

Mass balance in liquid: $\quad\quad\quad\quad \varepsilon \dfrac{\partial c_l}{\partial t} + \nabla \cdot \vec{N}_l = \dfrac{ai}{F} \quad\quad\quad$ (S6)

Butler-Volmer kinetics: $\quad i = i_0\left[\exp\left(\dfrac{\beta F \eta}{RT}\right) - \exp\left(-\dfrac{(1-\beta)F\eta}{RT}\right)\right] \quad$ (S7)

where $\vec{i}_s = -\sigma \nabla \phi_s$, $\vec{i}_l = -\kappa \nabla \phi_l + \dfrac{\kappa_D RT}{F}\nabla \ln c_l$, $\vec{N}_l = -D_l \nabla c_l + \dfrac{\vec{i}_l t_+^0}{F}$, and a Neumann boundary condition is used to relate the interfacial current density $i$ to $(\partial c_s / \partial r)\big|_{t,r_0}$, along with a no-flux condition at the centre of the active particles. Here, $\vec{i}_s$ and $\vec{i}_l$ are the solid-phase and liquid-phase current density, $a$ is the specific interfacial surface area within the porous electrode, and $i$ is the reaction current per unit area of the pore surface; $t$ is time, $r$ is radial position within the intercalation particle at a given location, $c_s$ is the molar concentration, and $D_s$ is the solid-phase diffusion coefficient of lithium in the intercalation particle; $F$ is Faraday's constant, $\varepsilon$ is the electrode porosity, $c_l$ is the liquid-phase salt

molarity, and $\vec{N}_1$ is the superficial salt flux. In the kinetic rate law, $i_0$ is the exchange-current density, $\beta$ is the symmetry factor, $\eta$ is the surface overpotential, $R$ is the universal gas constant and $T$ is temperature. In the flux laws, $\phi_s$ and $\phi_l$ are the solid-phase and liquid-phase electrical potentials, respectively; $\sigma$ and $\kappa$ are the effective electrical conductivity in the solid and liquid, respectively; $\kappa_D$ is the effective conductivity associated with concentration overpotential (formally related to transference number and thermodynamic factor), $D_l$ is the effective liquid-phase Fickian salt diffusivity, and $t_+^0$ is the cation transference number relative to the solvent velocity. Last, $r_0$ is the average radius of the intercalation particles. Note that in simulations, the $r$ direction is a pseudo-dimension orthogonal to the $x$, $y$, and $z$ directions (cf. Fig. S9 below) at every grid point, so that the model has four effective spatial dimensions.

When exchange-current density is large (or overpotential is small), the reaction kinetics can formally be linearized, yielding

$$i = i_0 \frac{F\eta}{RT}, \tag{S8}$$

a relation that is independent of the symmetry factor. If fast electrolyte diffusion is assumed in the liquid phase, then $D_l \to \infty$, and liquid-phase concentration polarization can be neglected, such that

$$\nabla c_l = 0. \tag{S9}$$

Accordingly, $\vec{i}_l \approx -\kappa \nabla \phi_l$, equation S9 becomes redundant, and the remaining balances and flux laws reduce to the Newman–Tobias porous-electrode model.[6] Rearranging the remaining balances in light of equations S8 and S9 gives the battery model we describe in the Experimental Procedures.

It is also possible to obtain the prior streamlined model of Chu et al.[1] within this framework. If lithium diffusion in the solid particles is assumed to be fast, then

$$\frac{\partial c_s}{\partial r} = 0 \quad \text{and} \quad \frac{\partial c_s}{\partial t} = -\frac{ai}{F}, \tag{S10}$$

where the second result owes to the divergence theorem. Thus, the governing system becomes

Charge balance in solid: $\quad\quad\quad\quad \nabla \cdot \vec{i}_s = -ai \quad\quad\quad\quad$ (S11)

Charge balance in liquid: $\quad\quad\quad\quad \nabla \cdot \vec{i}_l = ai \quad\quad\quad\quad$ (S12)

Mass balance in solid: $\quad\quad\quad\quad \dfrac{\partial \langle q \rangle}{\partial t} = -\dfrac{ai}{Fc_{s,\max}} \quad\quad\quad\quad$ (S13)

Linear reaction kinetics: $\quad\quad\quad\quad i = i_0 \dfrac{F\eta}{RT} \quad\quad\quad\quad$ (S14)

In which equation S13 replaces equation S5. Here $\langle q \rangle$ is a new variable that expresses the average SOC of the intercalation particle and $c_{s,\max}$ is the maximum lithium concentration in the solid.

## Supplemental Note 3.   Model simplification via scaling analysis

Chu et al.[1] identify a set of dimensionless parameters whose values can be rescaled to replace a fully resolved multiple-layer model with a homogenized single-layer model. If the dimensionless quantities describing the fully resolved model should remain identical to those describing the homogenized model, then the following rescaling of parameters is required:

$$\sigma_1 = N^2\sigma_N, \quad \kappa_1 = N^2\kappa_N, \quad \sigma_{1,cc} = N^2\sigma_{N,cc}, \quad \kappa_{1,sep} = N^2\kappa_{N,sep}, \tag{S15}$$

in which the subscripts "1" and "$N$" represent the properties of the (homogenized) single-layer and (fully resolved) $N$-layer cell, respectively.

These rescalings were incorporated into the COMSOL Multiphysics implementation. Apart from the battery geometry, copper bars used as heat sinks connected to the tabs were also included in the model geometry, as shown below in Fig. S9. The geometry was meshed and solved using COMSOL.

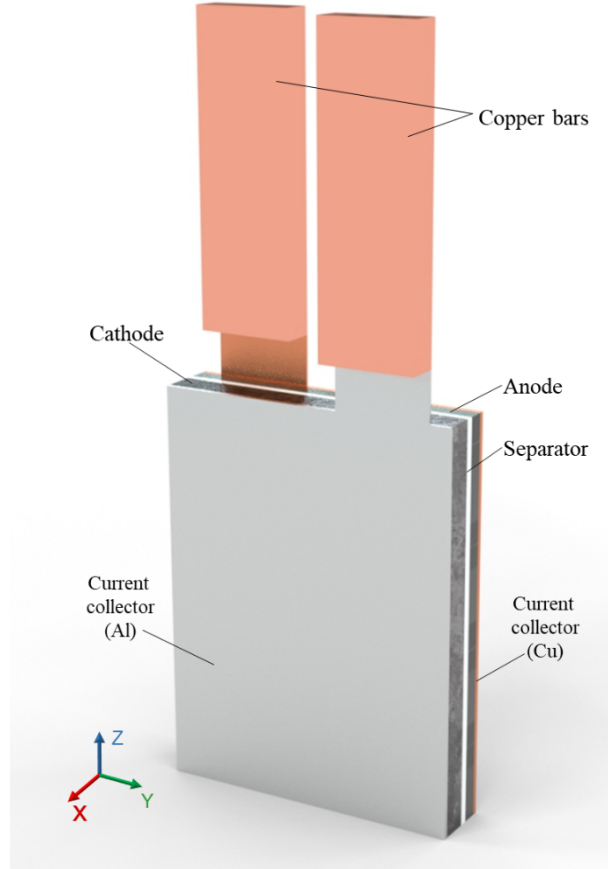

**Figure S9.** 3D single-layer battery geometry.

## Supplemental Note 4.  Model parameterization for full-discharge simulations

The pseudo-OCP of the LFP pouch cell was measured at 25 °C as a function of SOC by galvanostatic charge and discharge. Exemplary pseudo-OCP data are shown in Fig. S7. In simulations, this OCP was assigned to the LFP cathode, while the graphite OCP was taken to be ground, i.e., the lithiated graphite at its average SOC was taken to be the thermodynamic reference for cell OCP.

Manufacturer data for the temperature derivative of OCP, shown in Fig. S8, was used to obtain the SOC-dependent entropy change within the pouch cell, through

$$\Delta S(q) = F\left(\frac{\partial U}{\partial T}\right)_q. \tag{S16}$$

These data were validated using the entropy changes that resulted from fitting square-wave cycling tests at 30%, 50% and 70% SOC. After conversion to OCP derivatives using equation S16, these points were plotted on Fig. S8 to illustrate the good agreement between the fitted entropies and the

manufacturer data. In simulations, the entropy change due to the cell reaction was assumed to be divided equally between the positive and negative electrode.



\end{document}